# A LOW-LOSS FIBER ACCESSIBLE PLASMON PHOTONIC CRYSTAL WAVEGUIDE FOR PLANAR ENERGY GUIDING AND SENSING

Stefan A. Maier*, Paul E. Barclay, Thomas J. Johnson, Michelle D. Friedman, and Oskar J. Painter

*Thomas J. Watson Laboratory of Applied Physics, California Institute of Technology, Pasadena, CA 91125, USA*

**ABSTRACT**

A metal nanoparticle plasmon waveguide for electromagnetic energy transport utilizing dispersion engineering to dramatically increase lateral energy confinement via a two-dimensional pattern of Au dots on an optically thin Si membrane is described. Using finite-difference time-domain simulations and coupled-mode theory, we show that phase-matched evanescent excitation from conventional fiber tapers is possible with efficiencies > 90 % for realistic geometries. Energy loss in this waveguide is mainly due to material absorption, allowing for 1/e energy decay distances of about 2 mm for excitation at telecommunication frequencies. This concept can be extended to the visible regime and promises applications in optical energy guiding, optical sensing, and switching.

Corresponding Author:

Stefan A. Maier
Thomas J. Watson Laboratory of Applied Physics, California Institute of Technology
Mail Code 128-95, Pasadena, CA 91125
tel 626 395 6160
fax 626 795 7258
stmaier@caltech.edu

*electronic address: stmaier@caltech.edu

Research into electromagnetic surface modes or surface plasmon polaritons confined to interfaces between a metal and a dielectric has a rich history dating back to the mid-20th century [1, 2]. More recently, the emergence of sophisticated nanofabrication techniques has allowed fabrication of nanoscale metal patternings and led to intense research efforts due to potential applications in novel metal-based wavelength- and subwavelength-scale photonic devices [3]. One basic building block of such devices are plasmon waveguides that guide electromagnetic energy in the visible and near-infrared regime of the electromagnetic spectrum. Indeed, energy transport at telecommunication wavelengths using micron-size metal slabs has been demonstrated [4], and guiding at near-infrared and visible frequencies using nanowires [5] and arrays of closely spaced metal nanoparticles [6] has been achieved as a step towards the ultimate goal of building highly integrated photonic circuits [7] with spatial dimensions below the diffraction limit. These guiding structures may find practical applications, such as delivering energy to "hot-spots" for biological sensing [8], if two obstacles can be overcome.

The first of these issues concerns the difficulty of efficient excitation of localized plasmons in nanowires and nanoparticle structures. While plasmons on planar metal films can be efficiently excited using either prism or grating couplers [2], such methods cannot easily probe localized states in waveguides or resonators. Therefore, most investigations of localized states have employed either diffraction-limited far-field excitation with estimated efficiencies of about 15 % [9], or near-field optical fiber probes as local light sources with typical light throughputs well below 0.1 % [10]. Furthermore, neither of these methods is mode-selective. For metal slabs embedded in $SiO_2$ [4] and polymers [11], end-fire coupling has been demonstrated, but its efficiency is expected to decrease for asymmetric environments [12] necessary for surface sensing applications.

The second obstacle is the inherent trade-off between confinement and loss in plasmonic guiding structures. For metal slabs, it has been shown that the confinement of all supported long-ranging plasmon-polariton modes diminishes as the cross section of the waveguide (and thus losses due to ohmic resistive heating) are reduced [12]. Furthermore, asymmetry in the dielectric environment leads to a cut-off metal slab height for all long-ranging modes that increases both with dielectric contrast between the sub- and superstrate and decreasing width as nanowire dimensions are approached. For nanowires or one-dimensional nanoparticle waveguides, in order to reduce radiative loss, excitation near the plasmon resonance is necessary to provide the large lateral wave vectors required for a tight confinement to the submicron guiding structures. The requirement of resonant excitation increases the resistive heating losses, and further limits such waveguides to the visible regime of the spectrum. Therefore, typical 1/e energy attenuation distances are in the micron and sub-micron regime for nanowires and nanoparticle waveguides, respectively.

In this paper, we present the design of a two-dimensional metal nanoparticle plasmon waveguide that shows negligible radiative losses even for non-resonant excitation and analyze its guiding properties and excitation by conventional fiber tapers using a combination of finite-difference time-domain (FDTD) simulations and coupled-mode theory. Due to the employment of design principles developed initially for defect waveguides in dielectric photonic crystals, the energy attenuation length is mainly determined by the wavelength-dependent absorptive losses of the constituent metal nanoparticles. The confinement of the optical modes is ensured by a hybrid structure of SOI (silicon-on-insulator) and lithographically defined metal nanoparticles on the optically thin silicon membrane. Figure 1 shows this geometry both in top (b) and side (c) view for a plasmon waveguide based on a square lattice with a lattice constant of $\Lambda_z =$ 500 nm chosen to allow efficient excitation of the waveguide around a wavelength of 1.5 micron as discussed below. The metal dots are 50 nm in height, and have lateral dimensions of 100x100 $nm^2$ in the center of the waveguide with a linear lateral grading down to 70x70 $nm^2$ after three lateral periods. We chose gold with a dielectric function as described by Johnson and Christy [13] as the material system for our FDTD simulations.

Figure 1a shows the FDTD – calculated band structure of plasmon waves propagating in the z-direction for this geometry near the X – point of the first Brillouin zone (blue dots). Due to the periodicity of the structure in the propagation direction, a small band gap appears at the zone boundary, whose size can be adjusted by scaling the size of the metal nanoparticles. Another consequence of the periodicity is the characteristic zone-folding of the dispersion relation. This way, the upper plasmon band crosses both the light-line of silica and air, suggesting the possibility of phase-matched excitation of the waveguide by using silica fiber tapers [14] as discussed below. Note that while this waveguide was designed for operation in the near-infrared regime of the electromagnetic spectrum, simple adjustment of the lattice constant should allow guiding of energy at visible and other frequencies, although with commensurate increase in ohmic heating loss.

Figure 1b and 1c show the optical mode profile (electric field intensity) of the waveguide mode at phase-matching with the silica light line in top and side geometry.

Due to the lateral grading in dot size of the lattice [14], the mode is laterally well confined to the center of the waveguide akin to metal slab waveguides with a width/height ratio >> 1 [12]. Vertically, the confinement is ensured both by bound metal/air surface plasmons and the undercut geometry of the underlying silicon slab. In our FDTD model, the power flow through the boundaries of the simulation volume was analyzed, confirming good lateral waveguide confinement even for only three lateral grading periods of the lattice.

The energy attenuation in the direction of power flow is thus mainly determined by resistive ohmic heating losses in the metal. Since the lateral confinement of the guided mode is provided by the extended two-dimensional lattice structure, the metal nanoparticles can now be excited far below their individual plasmon resonances, thus minimizing the absorptive losses. Indeed, for operation around 1.6 micron our calculations based upon values of the dielectric properties of gold in the literature [13] predict an energy decay constant of $\alpha = 4.9$ $cm^{-1}$, corresponding to a 1/e energy attenuation length of about 2 mm. The attenuation length is significantly less than that predicted for plasmons propagating along continuous Au slab waveguides of the same thickness embedded in silica or in polymers [4, 11] due to the small amount of metal employed in our waveguide, and is expected to diminish even further with decreasing metal thickness. Note that the energy attenuation in actual fabricated devices will be higher than the predicted value based on resistive heating alone, due to scattering losses induced by the metal surface morphology and fabrication imperfections. For waveguides with a smaller lattice constant that operate in the visible regime of the spectrum, the absorptive losses and thus the energy decay will increase, which can be partially counteracted by a material change to silver. The waveguides can thus easily be adapted to desired applications such as low-loss energy guiding in the near-IR or short-distance sensing applications in the visible regime of the spectrum.

The high intensity of the electric field at the top surface of the waveguide due to the bound plasmon state makes this mode easily accessible optically. In particular, efficient overlap with a fundamental mode of a fiber taper placed at a gap $d$ on top of the waveguide as depicted in the inset of Fig. 2 is possible. Following the theory developed in [14] for dielectric photonic crystal waveguides, Figure 2 shows the power transfer characteristics between the taper and the plasmon waveguide versus gap height and coupler length. A high power transfer efficiency of 90 % over a distance of 70 lattice periods (35 μm) can be achieved for a gap of $d = 250$ nm. For dielectric photonic crystal waveguides, it has recently been shown experimentally that high power transfer at such small gaps can indeed be achieved [15]. Furthermore, the transverse coupling coefficient between the taper and plasmon waveguide mode is significantly larger than the theoretical energy attenuation $\alpha$, suggesting that high coupling efficiencies can be realized experimentally even in the presence of fabrication non-idealities.

Apart from allowing a highly efficient excitation of the waveguide using fiber tapers, the high electric field intensities at the metal-air surface of the particles also allow for other intriguing applications apart from energy guiding. Specifically, the high field intensities can serve as "hot-spots" for the attraction and subsequent sensing of biological molecules [16] or for enhancement of non-linear interactions for switching applications in a suitable adapted waveguide geometry after the coupling region. Thus, a new generation of wavelength and sub-wavelength scale photonic devices as proposed in [7] seems feasible.

In summary, we have shown that a hybrid SOI and metal nanoparticle waveguide structure based upon a laterally graded two-dimensional lattice of metal nanoparticles can serve as an efficient optical waveguide. The energy attenuation in these waveguides is limited only by resistive ohmic heating losses. Efficient, phase-matched excitation of the waveguide using tapers formed from standard silica optical fiber is shown to be possible, with theoretical power transfers of > 90 % in the near-infrared.

Figure captions:

Figure 1. 3D finite-difference time-domain simulation of a plasmon photonic crystal waveguide. a) Photonic band structure (blue dots) showing phase-matching of a waveguide mode with the silica light line. b) Electric field distribution of this mode in top view (Au metal dots outlined in white). c) Lateral distribution of the electric field for this mode through silicon membrane. d) Scanning electron micrograph of waveguide including end-mirrors fabricated from Au nano-dots on SOI, showing the lateral grade in Au dot size

Figure 2. Efficiency of contradirectional power transfer into a plasmon photonic crystal waveguide upon phase-matching with a silica fiber taper of diameter 1 μm placed at a gap *d* above the top of the waveguide as depicted in the inset. The contour plot shows the totally coupled power versus taper gap and coupling length in lattice constants on a linear color scale.

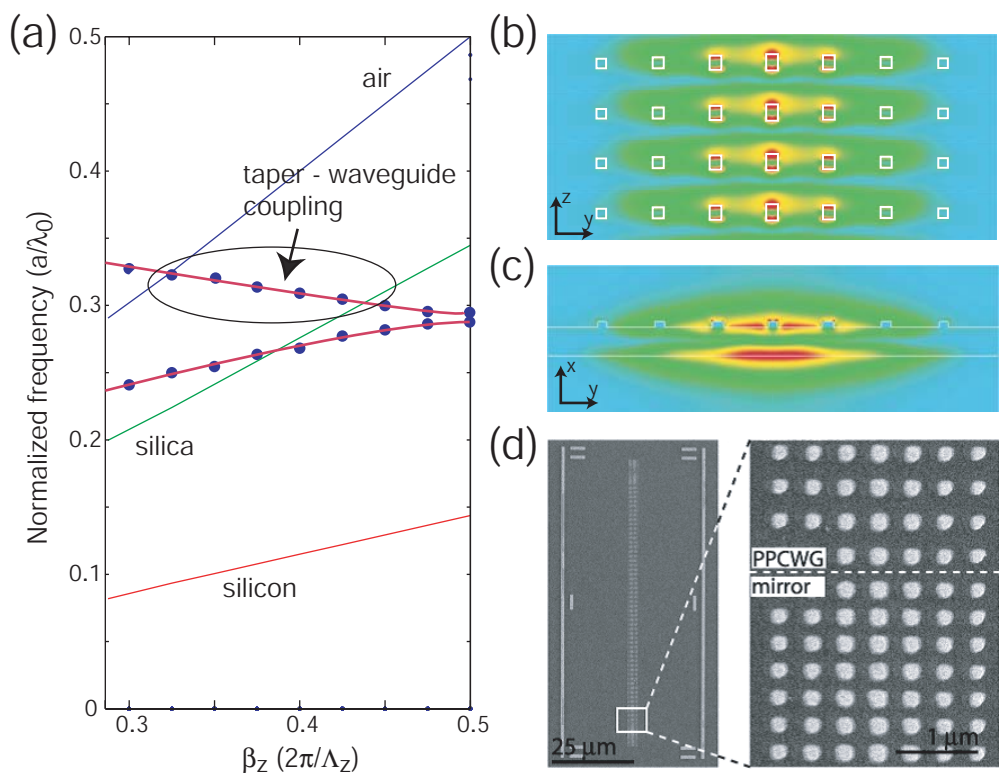

Figure 1 (color, double-column size) of 2
Stefan A. Maier, submitted to Applied Physics Letters

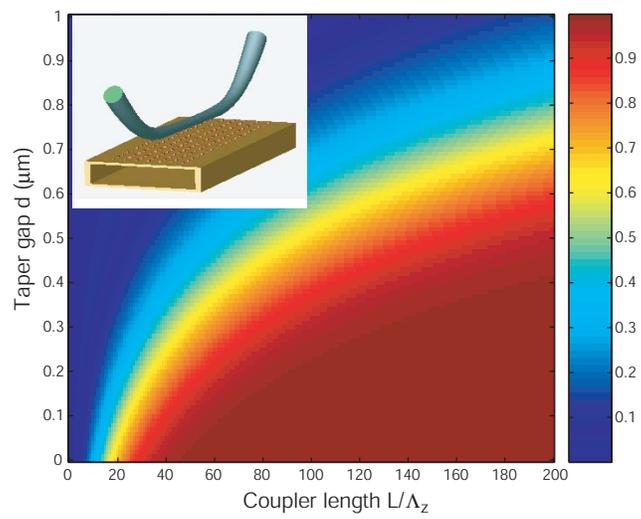